\documentclass[accept,preprints,article,moreauthors,pdftex]{Definitions/mdpi}
\firstpage{1} 
\pubvolume{10}
\issuenum{2}
\articlenumber{73}
\pubyear{2024}
\copyrightyear{2024}
\datereceived{14/12/2023} 
\daterevised{26/01/2024} 
\dateaccepted{28/01/2024} 
\datepublished{02/02/2024} 
\hreflink{https://doi.org/10.3390/universe10020073} 
\doinum{10.3390/universe10020073}
\pdfoutput=1 

\Title{Emergent Time and Time Travel in Quantum Physics}

\TitleCitation{Emergent Time and Time Travel in Quantum Physics}

\Author{Ana Alonso-Serrano $^{1,2,\dagger}$\orcidA, Sebastian Schuster~$^{3,}$*$^{,\dagger}$\orcidB{} and Matt Visser $^{4,\dagger}$\orcidC{}}

\AuthorNames{Ana Alonso-Serrano, Sebastian Schuster and Matt Visser}

\AuthorCitation{Alonso-Serrano, A.; Schuster, S.; Visser, M.}

\address{%
	$^{1}$ \quad Institut für Physik, Humboldt-Universität~zu~Berlin, Zum~Großen~Windkanal~6, 12489~Berlin, Germany\\
	$^{2}$ \quad Max-Planck-Institut für Gravitationsphysik (Albert-Einstein-Institut), Potsdam~Science~Park, Am~Mühlenberg~1, 14476~Potsdam, Germany; ana.alonso.serrano@aei.mpg.de\\
	$^{3}$ \quad Ústav Teoretické Fyziky, Matematicko-Fyzikální Fakulta, Univerzita~Karlova, V~Holešovičkách~747/2, 180~00~Praha~8, Czech~Republic; sebastian.schuster@utf.mff.cuni.cz\\
	$^{4}$ \quad School of Mathematics and Statistics, Victoria University of Wellington, P.O.~Box~600, Wellington~6140, New~Zealand; matt.visser@sms.vuw.ac.nz}

\corres{Correspondence: sebastian.schuster@utf.mff.cuni.cz}

\firstnote{These authors contributed equally to this work.} 

\abstract{Entertaining the possibility of time travel will invariably challenge dearly held concepts of fundamental physics. It becomes relatively easy to construct multiple logical contradictions using differing starting points from various well-established fields of physics. Sometimes, the interpretation is that only a full theory of quantum gravity will be able to settle these logical contradictions. Even then, it remains unclear if the multitude of problems could be overcome. Yet as definitive as this seems to the notion of time travel in physics, such a recourse to quantum gravity comes with its own, long-standing challenge to most of these counter-arguments to time travel: These arguments rely on time, while quantum gravity is (in)famously stuck with and dealing with the \emph{problem of time}. One attempt to answer this problem within the canonical framework resulted in the Page--Wootters formalism, and its recent gauge-theoretic re-interpretation—as an emergent notion of time. Herein, we will begin a programme to study toy models implementing the Hamiltonian constraint in quantum theory, with an aim towards understanding what an emergent notion of time can tell us about the (im)possibility of time travel.}


\usepackage{braket,mathrsfs,mathtools} 
\usepackage[locale = UK]{siunitx} 
\usepackage{dsfont} 
\usepackage{subcaption}
\usepackage[russian,newzealand]{babel}
\usepackage{csquotes} 

\allowdisplaybreaks 

\newcommand{\abs}[1]{\left\lvert #1 \right\rvert} 
\newcommand{\rbr}[1]{\left( #1 \right)} 
\newcommand{\sbr}[1]{\left[ #1 \right]} 
\newcommand{\oo}[1]{\frac{1}{#1}} 

\newcommand{\defi}{\mathrel{\mathop:}=} 
\newcommand{\ifed}{=\mathrel{\mathop:}} 
\newcommand{\tr}[1]{\ensuremath{\operatorname{tr}\left( #1 \right)}} 

\newcommand{\dif}{\ensuremath{\operatorname{d}}\!}

\newcommand{\HR}{\ensuremath{\hat{H}_{\textsc{r}}}} 
\newcommand{\HilR}{\ensuremath{\mathcal{H}_{\textsc{r}}}} 
\newcommand{\idR}{\ensuremath{\mathds{1}_{\textsc{r}}}} 
\newcommand{\omegaR}{\omega_{\textsc{r}}} 
\newcommand{\nR}{\hat{n}_{\textsc{r}}} 
\newcommand{\HC}{\ensuremath{\hat{H}_{\textsc{c}}}} 
\newcommand{\HilC}{\ensuremath{\mathcal{H}_{\textsc{c}}}} 
\newcommand{\idC}{\ensuremath{\mathds{1}_{\textsc{c}}}} 
\newcommand{\omegaC}{\omega_{\textsc{c}}} 
\newcommand{\nC}{\hat{n}_{\textsc{c}}} 
\newcommand{\ketastnull}{\ensuremath{\ket{\psi(0_\ast)}_{\textsc{c}}}} 
\newcommand{\ketasttheta}{\ensuremath{\ket{\psi(\theta_{\ast})}_{\textsc{c}}}}
\newcommand{\braasttheta}{\ensuremath{{\vphantom{\bra{\psi({{\theta}_{\ast}})}}}_{\textsc{c}}\! \bra{\psi({{\theta}_{\ast}})}}}

\begin{document}
	\section{Introduction}\label{sec:intro}
	As fascinating as the notion of time travel is, its long list of problematic issues—ranging from the classical~\cite{Roman:1992xj,Krasnikov:2001ci} \& \cite[p.212ff]{Visser:1995cc} to the quantum theoretical~\cite{Hawking:1991nk,Klinkhammer:1992tb,Visser:1992tx,Emparan:2021yon}—presents a very solid case against it, and its absence is effectively a feature of the universe accessible to human experience. Yet if one wants to look closely at these arguments, cracks and loopholes start to appear—many of which shall be mentioned in the following. To an extent, it almost seems as if the best evidence against any notion of time travel is that of our daily experienced notion of time and causality, our intuition. Still, the known universe is bigger than our day-to-day experience, and particularly the field of high energy physics is ripe with examples experimentally, observationally, or theoretically open to inquiry—in which various concepts or preconceived notions of human-scale physics have to be let go. In no field of theoretical physics is this more apparent than in the quest for quantum gravity. The field of contenders is large, and still growing: String theory, loop quantum gravity, quantum geometrodynamics, asymptotic safety, causal dynamical triangulation, causal set theory, Hořava--Lifshitz gravity, \dots Each comes with its own set of new notions beyond what can be experienced directly by human senses. Yet, these proposals still have to address, in one way or another, an issue plaguing quantum gravity from its beginning: Time. This \emph{problem of time}, and its eventual resolution, then naturally will also impact what these theories have to say about (or against) time travel.
		
	The \enquote{problem of time} in \emph{classical} canonical gravity is settled by a careful, gauge-theoretic analysis~\cite{Pons:2010ad} that distinguishes the different rôles played by the Hamiltonian, either in the constraint/gauge fixing or in the definition of observables, respectively. In quantum gravity, however, the \emph{problem of time} remains (at the very least) a topic of active discussion and research~\cite{Anderson:2017jij,DiGioia:2019nti,Kiefer:2021zdq}. The reason for this goes back even before a theory of quantum gravity was the focus: While there are ways to give position and momentum a meaning in terms of operators on (rigged) Hilbert spaces in quantum mechanics, this cannot be carried over to time and energy. This was concerning to researchers in quantum mechanics early on, as the time-energy uncertainty relations proved to be both reliable and instructive, while not fitting into the standard paradigm of the Robertson--Schrödinger uncertainty relations for (generic) Hermitian operators. Only the proof of Mandelstam and Tamm in 1945 improved this uncertainty relation's foundation, while begging the question of why a different approach was needed~\cite{MandelstamTamm1945TimeEnergy}. Essentially, a series of impossibility results for time operators canonically conjugate to a Hamiltonian was presented: Schrödinger used an approach based on normalizability~\cite{Schroedinger:1931SRuQM}, similar to the argument for why plane waves require a rigged Hilbert space. Pauli, in a footnote, used a simpler argument juxtaposing discrete spectra and the continuous translation in energy such a time operator would induce~\cite{Pauli:1990,Pauli:1980}. Finally (and in reply to what will be discussed shortly), Unruh and Wald gave a general argument for semi-bounded Hamiltonians linked to unitarity~\cite{Unruh:1989db}. Two at first glance quite distinct approaches were developed in reply to these no-go results.
	
	One approach was similar in spirit to the answer to early counterarguments to plane waves, \emph{e.g.}, eigenstates of definite position and momentum. While with position and momentum one relaxes the idea of being restricted to Hilbert space and instead works in a rigged Hilbert space~\cite{ABohmQM,delaMadrid:2001wwa,Thiemann:2007pyv}, here, when faced with the above-cited no-go results concerning time operators, one relaxes the idea of Hermiticity. Instead, one introduces the notion of a positive, operator-valued measure (POVM). It can represent not just perfect measurements yielding a definite eigenstate, but also imprecise measurements~\cite{BLPY2016QuantumMeasurement}. Besides allowing one to address the measurement problem in quantum field theory~\cite{Fewster:2018qbm}, phase observables~\cite{BGM1995POVMPhaseObservables}, or the already mentioned, imprecise measurement processes~\cite{AulettaFortunatoParisi2009QM,BLPY2016QuantumMeasurement}—and more important for the present context—this allows a notion of time measurement~\cite{BGL1994TimeObservables}. Since the no-go theorems relate to Hermitian operators, the more general POVMs are not ruled out as time observables. As we will see, the context of time travel allows one to avoid these no-go results by other means, too. Still, the bigger picture of gauge theory provides the most pertinent background for extensions of the model to be described in this paper.
	
	Another approach that developed independently from POVMs was to carefully distinguish between a clock and time. Paraphrasing Einstein~\cite{Einstein:1905Edyn}, time is meaningful only after specifying the clock measuring it. Early on, in developing canonical quantum gravity, DeWitt~\cite{DeWitt:1967yk} pointed out how different subsystem's operators in a given Hamiltonian constraint could serve this purpose of an effective clock. Page and Wootters then took this idea and made it more general and workable, by phrasing time evolution in terms of conditional probabilities linking different subsystems~\cite{Page:1983uc,Wootters1984Time}.\footnote{With a bit of hindsight, already Schrödinger anticipated this in his footnote on page~245 of~\cite{Schroedinger:1931SRuQM} [here, in our translation and keeping the somewhat convoluted grammar of the original]: \enquote*{An interesting application of this is the following: if one knows of a system composed of several, coupled subsystems only the total energy, then it is impossible to know more about the distribution of energy across the subsystems than the statistical, time-independent data, which already follows from the knowledge of the total energy. Except for the case that individual subsystems are in truth fully decoupled, energetically isolated from the others.}} This Page--Wootters (PW) formalism was soon challenged on various grounds by Unruh, Wald~\cite{Unruh:1989db} and Kuchař~\cite{Kuchar:1991qf}, leading to this approach lying dormant for some time.
	
	Recent developments then made the latter approach re-emerge~\cite{Giovannetti:2015qha,Marletto:2016gwv,Diaz:2018uny,Diaz:2019xie,Diaz:2020dfe,Gemsheim:2023izg}, and both approaches converge~\cite{Smith:2017pwx,Hoehn:2019fsy}. This also allowed one to address and re-contextualize the earlier criticisms of the PW formalism~\cite{Hoehn:2019fsy,Hoehn:2020epv}. In essence, the clocks chosen within the PW formalism become part of a gauge-theoretic picture of clocks; the \enquote{physical Hilbert space} being a \enquote{clock-neutral} picture, and different clocks representing different \enquote{gauge conditions}~\cite{Hoehn:2019fsy}. These results form part of a larger research programme that aims to ground the rulers and clocks at least implicitly employed in external symmetries with an operational underpinning~\cite{Krumm:2020fws,Goeller:2022rsx,Adlam:2022zar,Hoehn:2023axh,Hausmann:2023jpn,Hoehn:2023ehz}. For our purposes below, the key takeaway is that the PW formalism still works, if its defects and the criticisms of it are understood as related to issues already familiar from electromagnetism: One has to choose whether a calculation should be gauge-fixed or gauge-independent. Depending on the question at hand, one or the other will be better suited to calculations or understanding. Our model below will be too simple to fully showcase these developments, but given the criticisms the PW formalism faced, it is important to keep the resolution in mind.
	
	Within this special issue's scope, many other pertinent links of time and time travel to quantum physics could be made. A selected, non-exhaustive list would include: Deutsch CTCs and (non/retro-)causal quantum processes~\cite{Deutsch:1991nm,Bishop:2020qtt,Baumann:2021urf,Venkatesh:2020qvl,Adlam:2022uqw}, simulation of time (travel) in analogues~\cite{Jannes:2009ns,Barcelo:2022jbr,Sabin:2022mmj}, semi-classical/quantum stability~\cite{Hawking:1991nk,Klinkhammer:1992tb,Visser:1992tx,Visser:1997gq,Emparan:2021yon}, \emph{et cetera}. For the purposes of this article, however, we would like to highlight only two further research avenues. One is that of different \emph{notions} of time. Yet another, non-exhaustive list would include at least cosmological time, psychological time, parameter time, thermodynamic time, dynamical \emph{versus} kinematical time, \dots While this distinction is studied mostly in the context of \enquote{the arrow of time}~\cite{Zeh2007Time}, it is worth pointing out that many counterarguments to time travel have to rely at least on some conflation of these concepts. To give just one example, if thermodynamic arguments are brought forth, one needs to establish a more or less direct relation between the thermodynamic notion implied by the second law\footnote{One would also have to brush aside that standard thermodynamics has to artificially graft time onto its formalism in the first place, leading to the very name \enquote{thermo\emph{dynamics}} being rather unintuitive compared to other occurrences of \enquote{dynamics} in physical terminology~\cite{Truesdell:1971}.} and the notion of time invoked in time travel. (This itself does not have to be the same notion as that of general relativity (GR), where time travel can be identified with closed, time-like curves (CTCs) or related concepts~\cite{Minguzzi:2006sa}. In theories other than GR, a different notion of time might be relevant to describe time travel!) 
 
	The other important point to make is that despite all the counterarguments, time travel is just one notion of many with questionable \enquote{physicality}~\cite{Schuster:2023jfa}. Even within the context of GR, other notions of physicality vie for validity with the absence of time travel, for example, various kinds of completeness (geodesic, hole free, \dots) or the validity of various kinds of energy conditions. These notions, however, are not all compatible with each other~\cite{Barcelo:2002bv,Earman:2009mda,Manchak2009TimeMachines,Manchak:2018hut,Manchak2021CollectingCollections,Santiago:2021aup}. One might find oneself in the uncomfortable situation that a dearly held and important property (stability, completeness, fulfilled energy conditions, \dots) will require time travel to be permitted; and this just within GR. Other theories are unlikely to be free from such problems, especially as many of the comforting no-go theorems are specifically proven within the context of GR. It it therefore not surprising that in the context of classical field theories of gravity alone, there exist many different ways to realize time travel. This includes wormholes~\cite{Visser:1995cc,Gonzalez-Diaz:2011wel}, warp drives~\cite{Everett:1995nn}, cosmologies~\cite{Godel:1949ga,Hawking:1973uf,Griffiths:2009dfa,Gray:2016pbu}, maximally extended space-times (such as Kerr near the central ring singularity) \cite{Wald:1984rg}, and various more mathematical construction techniques~\cite{Penrose:1972,Krasnikov:1995ad,Manchak2011NoNoGo}. The latter can also be employed to distinguish between, for example, time travel (CTCs) and time machines (whatever structure creates or causes CTCs) \cite{Earman:2009mda,Manchak2009TimeMachines,Manchak2011NoNoGo}.
	
	This finally brings us to the goal of this paper: We want to study what can be said about the viability of time travel if time itself is only an emergent concept, as in the PW formalism. Concretely, we will be studying an example of two non-interacting harmonic oscillators similarly constrained  to a fixed total energy of zero as in a Wheeler--DeWitt (WDW) equation, \emph{i.e.}, it will mimic a minisuperspace model of time travel. This toy model is by no means meant to exhibit an exhaustive description of whether and how time travel can arise in the PW formalism. Nor should this model be taken literally as a minisuperspace model, as no gravitational model is canonically quantized to arrive at it; rather it demonstrates possible phenomenology. We will see that the results in our toy model point to an extraordinarily bland version of Novikov's self-consistency conjecture~\cite{Novikov:1989sd,Sklar:1990,Friedman:1990xc}.
	
	\subsection{Outline}
	Following the preceding introduction to and overview of the related research question in section~\ref{sec:intro}, we will provide an introduction into the methods employed: In section~\ref{sec:PW}, we briefly summarize the PW formalism. In section~\ref{sec:POVM} we demonstrate POVM's utility for implementing time observables in the context of a single harmonic oscillator. The main part of the paper is employing these concepts then, in section~\ref{sec:WDW}, to two harmonic oscillators subjected to a WDW-like Hamiltonian constraint equation. We close by discussing our results and pointing out future directions and open questions in section~\ref{sec:final}. Finally, the appendices collect additional details and context: Appendix~\ref{app:defPOVM} reviews the mathematical definition of a POVM, appendix~\ref{app:HO} collects results concerning periodic time in the quantum harmonic oscillator that supplement the discussion of section~\ref{sec:HO}, while appendix~\ref{app:fig} gives additional details concerning the construction of the paper's figures.

	We will use natural units in which $G=\hbar=c=1$.
	
	\section{The Page--Wootters Formalism}\label{sec:PW}
	Let us consider a quantum system (\enquote{the universe}) that separates into two subsystems, a \enquote{clock} system and a \enquote{residual} system. The PW formalism is our first, and most important ingredient for studying time travel without time (or rather: time travel with only an emergent notion of time). It is a proposal to give an operational meaning to measuring the time evolution of the residual subsystem in the larger quantum system (the \enquote{universe}) with respect to the other, (mostly) decoupled \enquote{clock} subsystem. Partly, this separation into \enquote{clock} and \enquote{residual} (in the literature often \enquote{rest}) systems is motivated by the problem of time encountered in quantum gravity. In terms of the total Hilbert space $\mathcal{H}$, the clock Hilbert space $\HilC$, and $\HilR$, the residual Hilbert space, this means:
	\begin{equation}
		\mathcal{H} = \HilC \otimes \HilR.\label{eq:HilSep}
	\end{equation}
	Subsequent developments and refinements sometimes further separated what we call $\HilR$ into evolving and ancillary/memory/supplementary/\dots systems, but for our purposes, we will not follow this line of thinking—at least in this current paper. In future work, it could be interesting to see whether such an approach could provide a notion of time travel that—contrary to GR's CTCs—only allows for a finite number of temporal round trips. Each of these Hilbert spaces is equipped with a Hamiltonian acting on these separate Hilbert spaces, such that 
	\begin{equation}
		\hat{H} \ket{\Psi} = \rbr{\HC \otimes \idR + \idC \otimes \HR} \ket{\Psi} = E \ket{\Psi},\label{eq:GenHamiltonian}
	\end{equation}
	where $\ket{\Psi}$ is a state in the total Hilbert space $\mathcal{H}$.	In line with Schrödinger's 1931 footnote, this state $\ket{\Psi}$ is required to be in a definite, fixed eigenstate of $\hat{H}$ with total energy $E$. More generally, the PW formalism remains viable even for density matrices $\hat{\rho}$, with the above case regained by setting $\hat{\rho} = \ket{\Psi}\bra{\Psi}$.
	
	Examples of such systems usually are constructed by symmetry-reducing the WDW equation of quantum gravity, the quantized Hamiltonian constraint of GR:
	\begin{equation}
		\hat{H} \ket{\Psi} = 0.\label{eq:WdWgen}
	\end{equation}
	In this expression, we gloss over many technicalities (such as operator ordering, domain issues, topological concerns, configuration space considerations, \dots) \cite{DeWitt:1967yk,Thiemann:2007pyv,Kiefer:2012book}, as our focus lies less on quantum gravity and more on the emergent notion of time that the PW formalism provides. Still, we will occasionally take inspiration from quantum gravity. In particular, one can consider our final model to be a crude example of a minisuperspace model that has \emph{not} been found through symmetry reduction. Already here, we therefore want to point out that many questions the minisuperspace models of quantum black holes or quantum cosmology try to answer, will not and cannot appear in our context. Most notably will be an absence of gravitational singularities, whose avoidance has been a mainstay of quantum gravity.
	
	The key step for formulating the PW formalism now lies in describing the evolution of the system encoded through a tuple $(\HilR,\HR)$ in terms of the evolution of $(\HilC,\HC)$. For this, one picks some initial state $\ketastnull \in \HilC$, which is then evolved using the clock Hamiltonian according to
	\begin{equation}
		\ketasttheta \defi e^{-i\HC \theta_{\ast}} \ketastnull. \label{eq:ketasttheta}
	\end{equation}
	This corresponds to the familiar idea of unitary time evolution of states in the Schrödinger picture through a Hamiltonian, but—at this level of the discussion, anyway—is a definition. The Schrödinger or Heisenberg equations of motion would then be a \emph{result} of this approach~\cite{Page:1983uc,Smith:2017pwx}.
	
	The evolution of the subsystem $\HR$ is then evaluated in terms of relative probabilities with respect to these clock states. To do this, let us first define the projector
	\begin{equation}
		\hat{P}_{\theta_{\ast}} \defi \rbr{ \ketasttheta\ \braasttheta } \otimes \idR \label{eq:Ptheta}
	\end{equation}
	of a given state onto particular \enquote{clock time states} $\ketasttheta$ in $\HilC$, leaving states in $\HilR$ unchanged. Then the evolution in $\HilR$ by $\HR$ with respect to clock states is given by the expression
	\begin{equation}
		E(\HR|\theta_{\ast}) = \frac{\tr{\HR \hat{P}_{\theta_{\ast}} \hat{\rho}}}{\tr{\hat{P}_{\theta_{\ast}} \hat{\rho}}}.\label{eq:CondProb}
	\end{equation}
	This gives the conditional expectation value of the residual Hamiltonian $\HR$ when the clock (state) \enquote{reads} $\theta_{\ast}$. The technical requirements behind this interpretation are stationarity, 
	\begin{equation}
		[\hat{H},\hat{\rho}]=0,
	\end{equation}
	and commutativity of $\HR$ and $\HC$, which we enforced by the block structure of $\hat{H}$ encoded in equation~\eqref{eq:GenHamiltonian}. These assumptions are quite strong, especially once one wants to include an interaction Hamiltonian coupling $\HilC$ and $\HilR$ on the right hand side of equation~\eqref{eq:GenHamiltonian}. Given the recent clarification on this issue provided by~\cite{Giovannetti:2015qha,Marletto:2016gwv,Smith:2017pwx,Hoehn:2019fsy}, we want to return to such generalizations at a later point, but stick to these more restrictive assumptions in the present context. 
	
	Some general comments are appropriate at this point: First, fixing the initial, \emph{total} state $\ket{\Psi}$ can, in general, greatly change the time evolution as described by equation~\eqref{eq:CondProb}. Second, to reiterate: This is not the most sophisticated or modern implementation of the PW formalism, see the previous paragraph. Third, without further specifying a particular system comprised of the various Hamiltonians present in equation~\eqref{eq:GenHamiltonian}, we cannot even begin our investigation. This holds true for the sort of time evolution people expect—one without time travel—just as well as for the more unexpected time evolution—showcasing a notion of time travel. The next step will be to lay the groundwork for our toy model, specifying the state and the evolution.

	\subsection{Time Travel in the Page--Wootters Formalism}
	In order to prepare this next step, let us first connect the PW formalism to some of the notions mentioned in the introduction. In particular, we would like to be able to talk about time travel in this setting. The simplest way to achieve this is to demand that the domain from which our time parameter $\theta$ is taken to be $S^1$, \emph{i.e.}, periodic. By itself, this will not guarantee a useful notion of time travel. A simple way to see this is by looking at everyday clocks: They are all periodic. Only a \emph{calendar}, also keeping track of years, would be a truly non-periodic clock. One does not travel back in time when one wakes up at the same time each morning, just because the alarm clock is periodic.
	
	Instead, we will try to find a way to implement the simplest possible version of time travel. This is (Novikov) self-consistency. When the clock returns to a state, so will the remainder system return to the same state it had when the clock last was in this state. This situation can (at least in some systems and models) also yield existence results, though usually at the expense of uniqueness~\cite{Friedman:1990xc,Echeverria:1991nk,Fewster:1996tt,Bachelot2001GlobalWEchrono,Dolansky:2010nr}. So far, however, the PW formalism as presented here does not allow for this. The evolution of the clock as prescribed by equation~\eqref{eq:ketasttheta} is ignorant of the domain from which to draw $\theta_\ast$.
	
	In order to alleviate this problem, we will borrow concepts from the POVM approach to time operators already mentioned in the introduction, and to be discussed in more detail in the next section. In order to incorporate this in the PW formalism, we note that the key ingredient for comparing the evolution of the clock with that of the remainder system is encoded in the definition of the projector onto clock states, equation~\eqref{eq:Ptheta}. The goal, thus, will be to not only influence the clock evolution through a convenient choice of the global state $\ket{\Psi}$. Rather, we also will have to make a judicious choice of the clock states themselves, and how to identify them in a meaningful way.
	
	\section{POVMs, the Harmonic Oscillator, and Time}\label{sec:POVM}
	The second ingredient of our investigation is a particularly simple example of time operators and POVMs: The harmonic oscillator~\cite{Susskind:1964,Garrison:1970,Galindo:1984,BGL1994TimeObservables,BGM1995POVMPhaseObservables}. As an ubiquitous model system in all fields of classical and quantum physics, it is likely to be an instructive ingredient even in our systems following a Hamiltonian constraint similar to the WDW equation~\eqref{eq:WdWgen}. Our discussion will follow the presentation of the two just-mentioned articles, adapted to the notation chosen for our application. Concretely, it is possible in the POVM framework to introduce time with a periodic domain through a phase POVM. Similar to time operators, also an operator representing a phase variable has been difficult to construct in a mathematically rigorous manner. Here, the problems are tied to issues of boundary conditions, hermiticity, and the domain of operators not given in standard notation. As with time operators, the more general setting of POVMs provides a way around this. Since a time operator ideally would be canonically conjugate to the Hamiltonian (as suggested by the energy-time uncertainty relation), and a phase operator ideally would be canonically conjugate to the number operator, the connection to periodic time is easily made.

	In our toy model of harmonic oscillators (further specified below in section~\ref{sec:WDW}), we will identify phase and time. This is also provides the easiest way to rigorously define time travel, making use of Novikov's self-consistency conjecture: A system comprised of clock and dynamical system with respect to clock time undergoes time travel, if the dynamical system state is the same after each full clock period. This definition could, in principle, be relaxed. For example, this could happen only a finite number of times. An emergent notion of time furthermore allows one to make such a statement merely approximate. This will allow us to study rigorously notions of time travel that not fully fall into this framework of self-consistency; for example, only parts of the system could undergo time travel. We hope that such models including the clock itself can shed light on when a system is \enquote{merely periodic in time} \emph{versus} when it actually undergoes time travel. In our final section~\ref{sec:final}, we will further expound on this. Importantly, a requirement of self-consistency will preempt any of the traditional paradoxa constructed out of it. 

	\subsection{Time and Phase Operators for the Harmonic Oscillator}\label{sec:HO}
	Leaving the mathematical formalities (mostly) aside, we will now introduce the specific POVM used in our simple model of time travel. Formal, supplementary details regarding the definition of POVMs can be found in appendix~\ref{app:defPOVM}. For the purposes of this section, we will drop all label indices, and only look at the simple, standard harmonic oscillator with Hamiltonian
	\begin{equation}
		\hat{H} = \oo{2} \hat{a}^\dagger \hat{a} + \oo{2}. \label{eq:HO}
	\end{equation}
	We want to construct a time operator for this. In order to do so, we look at the \emph{polar decomposition} of the ladder operator $\hat{a}$, \cite[Thm.~VIII.32]{Reed:1980}:
	\begin{equation}
		\hat{a} = \hat{W} \widehat{\abs{a}}.
	\end{equation}
	Here, $\widehat{\abs{a}}$ is easy to define as
	\begin{equation}
		\widehat{\abs{a}} \defi \hat{n}^{1/2} = \sqrt{\hat{a}^\dagger \hat{a}}.
	\end{equation}
	The \enquote{operator} $\hat{W}$, however, is less simple than its occurrence in a \enquote{polar decomposition} might suggest: It will not be unitary. More concretely~\cite{BGM1995POVMPhaseObservables}:
	\begin{equation}
		\hat{W} \hat{W}^\dagger = \mathds{1},
	\end{equation}
	but
	\begin{equation}
		\hat{W}^\dagger \hat{W} = \mathds{1} - \ket{0}\negthickspace\bra{0} \neq \mathds{1}.
	\end{equation}
	
	The phase or time states to be used are now constructed as eigenstates $\ket{\theta}$ of $\hat{W}$. 
	\begin{equation}
		\hat{W}\ket{\theta} = e^{i\theta} \ket{\theta}.\label{eq:clockstate}
	\end{equation}
	While these eigenstates do exist, given the properties of $\hat{W}$, they are now only an \emph{over}complete set of eigenstates.\footnote{Such overcomplete states find ample application in, for example, the context of coherent states~\cite{AulettaFortunatoParisi2009QM}.} So different eigenstates $\ket{\theta}$ and $\ket{\theta'}$ will not be orthogonal, and expressing a general state $\ket{\psi}$ in terms of these eigenstates will not be unique anymore. Each of these states $\ket{\theta}$ can be written as
	\begin{equation}
		\ket{\theta} = \sum_{n\geq 0} e^{in\theta} \ket{n}.\label{eq:clockstate-fock}
	\end{equation}
	
	Now is the time to define a concrete POVM. In this instance, let us define for all Borel sets $X$ of $[0,2\pi)$
	\begin{align}
		B_0(X) &\defi \oo{2\pi} \int_X \ket{\theta}\negthickspace \bra{\theta} \dif \theta,\label{eq:B0}\\
		&= \sum_{n,m\geq 0} \oo{2\pi} \int_X e^{i(n-m)\theta} \ket{n}\negthickspace\bra{m} \dif \theta.
	\end{align}
	The index $0$ is a first indication of what is to come: Different points on the unit circle $S^1$ can be chosen as starting points. In the above definition, one starts on the positive real axis, \emph{i.e.}, we choose our angle $\theta$ from $[0,2\pi)$, as opposed to say from $[\theta_{\ast}, \theta_{\ast}+2\pi)$ for some $\theta_{\ast} \in [0,2\pi)$.
	
	Using
	\begin{equation}
		\oo{2\pi}\int_o^{2\pi} e^{i(n-m)\theta} \dif \theta  = \delta_{nm}
	\end{equation}
	and
	\begin{equation}
		\sum_{n=0}^\infty \ket{n}\negthickspace\bra{n} = \mathds{1}
	\end{equation}
	one can show that
	\begin{equation}
		B_0([0,2\pi)) = \mathds{1},
	\end{equation}
	which means that $B_0$ is a normalized POVM. So $B_0$ represents an observable in the sense of POVMs. Note that in the context of the definition of a POVM, in the present example we have that $\Omega=S^1$.
	
	From the POVM side of things, this gives all ingredients needed for the calculations to follow. However, for a more comprehensive understanding of the physics of time involved in the harmonic oscillator, it is worthwhile to point out additional considerations concerning such periodic time variables. We have collected these observations in appendix~\ref{app:HO}.

	\section{Time Travel in Two Harmonic Oscillators Subject to a Hamiltonian Constraint}\label{sec:WDW}
	The goals for the present article are modest. We want to study as simple a model as possible that is subject to a constraint reminiscent of the WDW equation~\eqref{eq:WdWgen}. As we also want to capture time travel as defined above, we need for this two ingredients: A periodic clock system $(\HilC,\HC)$ and a system $(\HilR,\HR)$ that evolves in the sense of the PW formalism (at least) as periodic as the clock itself. Then it would be an instance of a quantum system fulfilling the Novikov self-consistency conjecture. Based on the discussion of harmonic oscillators above, the simplest such system would be to use a single harmonic oscillator for each $\HR$ and $\HC$, respectively. To allow for some more generality, we will not demand their respective frequencies $\omegaR$ and $\omegaC$ to be the same, \emph{a priori.} Then, our model will have the following Hamiltonian in the full Hilbert space $\mathcal{H}$:
	\begin{equation}
		\hat{H} \ket{\Psi} = \rbr{\vphantom{\omegaC \,\hat{n} + \omegaC \oo{2}\idC}
			\smash{\underbrace{\omegaC \,\nC + \frac{\omegaC}{2}\,\idC}_{\HC} \underbrace{- \omegaR \,\nR
					-\frac{\omegaR}{2}\, \idR}_{-\HR}}}\ket{\Psi}=0.
		\vphantom{\underbrace{\omegaC \hat{n} + \omegaC \oo{2}\idC}_{\HC}}	 
		\label{eq:WdW}
	\end{equation}
	This means that the system Hamiltonian $\HR$ and the clock Hamiltonian $\HC$ are part of a WDW-like equation. Naturally, this implies that it falls into the framework of the PW formalism for a stationary Hamilton equation $\hat{H} \ket{\Psi} = E\ket{\Psi}$ with energy $E=0$.
	
	The previous two sections gave us the two most important ingredients in place: First, the way to measure time evolution with respect to a clock system using the PW formalism's conditional probability~\eqref{eq:CondProb}. Second, the required clock states given by equation~\eqref{eq:clockstate} with respect to which we will measure evolution, and its representation in terms of the harmonic oscillator eigenstates, equation~\eqref{eq:clockstate-fock}. Now, we can make use of it by implementing the Hamiltonian constraint~\eqref{eq:WdW} into this model. What remains to be done, though, is to specify more explicitly the type of total state $\ket{\Psi}\in\mathcal{H}$ the \enquote{universe} (as described by $\hat{H}$) is in.
	
	\subsection{Important Note on the Difference to Quantum Cosmology}
	Having introduced the model, it is easy to recognize it as a variant of a model already used in the context of quantum cosmology~\cite{Kiefer:1989km,Kiefer:1989va,Page:1990mh,Brunetti:2009eq}. There is, however, an important distinction both mathematically and interpretationally. One of the goals of quantum cosmology is to use symmetry reduction to arrive at minisuperspace models testing aspects of quantum gravity. In the context of (quantum) cosmology, this is in particular often the status of Hawking-type singularity theorems in quantum gravity: Can a cosmological singularity be avoided or not~\cite{Hawking:1965cc,DeWitt:1967yk,ONeill:1983,Kiefer:2012book}? In many, if not most situations this leads to the demand that the configuration space $\mathds{R}^2$ of the two harmonic oscillators is limited to the half-plane $\mathds{H}^2$, as one of the variables corresponds to the non-negative scale factor $a$. The scale factor at $a=0$ corresponds to a gravitational singularity. Restricting one harmonic oscillator to only a half-plane for its configuration space variable, however, inevitably leads to the inability to use the very convenient, powerful, and simple algebraic methods in terms of ladder operators used so far in the present paper.
	
	Our present goals, however, are different. We want to study time travel in a quantum-theoretic framework, but in a manner that is as theory-agnostic as possible. Hence, we will simply not assume that equation~\eqref{eq:WdW} necessarily has its origins in a Hamiltonian formulation of general relativity or similar gravitational theories. This means that equation~\eqref{eq:WdW} with the domain used is an explicit input. It very well could be derived from a specific gravitational model in a specific gravitational theory through symmetry-reduction. Should this not be possible, it still may serve as a toy model. Especially, as the process of symmetry-reduction (to arrive at minisuperspace models) itself may be a fraught undertaking. Our assumption will be that such an equation with the \emph{full} configuration space $\mathds{R}^2$ \emph{is} available, and, likewise, so are our corresponding trusty ladder operators. This also bypasses issues such as the precise nature of the inner product on the quantum gravitational wavefunctions, as already discussed in \cite[\S9]{DeWitt:1967yk}. Two extremal options for inner products (with many in between) are a quantum harmonic oscillator on the one side and a wave equation on the other. We assume the inner product structure as the one familiar from and resulting in standard quantum mechanics. Naturally, this leads to many directions into which to extend our present model, and we will come back to these in our concluding section~\ref{sec:final}.
	
	In conclusion and in conscious distinction to quantum cosmology, our toy model~\eqref{eq:WdW} for time travel will be treated using a ladder operator approach of quantum mechanics. This allows us to use the methods introduced in the previous two section, and hopefully also provides easy extensions in diverse directions.

	\subsection{A Boring Model of Time Travel}
	Looking for a square-integrable solution to equation~\eqref{eq:WdW} introduces a new, additional condition on this equation. Only if
	\begin{equation}
		\frac{\omegaR}{\omegaC} = \frac{2 n_{\textsc{c}} + 1}{2 n_{\textsc{r}} + 1},\qquad \text{where~}n_{\textsc{c}},n_{\textsc{r}} \in \mathds{N}_0,\label{eq:commensurability}
	\end{equation}
	will this be possible. This \emph{commensurability condition} will have far-reaching consequences for our simple toy model in what is to come.\footnote{Deviating from the commensurability condition even a little would lead to non-normalizable wave functions. In the context of quantum cosmology such wave functions can be encountered, such as the no-boundary proposal \cite[p.282]{Kiefer:2012book}. For the sake of our (not necessarily gravitational) toy model, we believe that this would add another layer of interpretational problems best left for a separate study.} In fact, these results will (for this toy model, at least) be so far reaching, that the precise nature of the time operators employed or their respective states used as initial state of the clock becomes immaterial. The interested reader can find more details on a possible choice of time operators useful for the current purposes in appendix~\ref{app:HO}.
	
	Quite generally, the  wave function of the \enquote{universe}, $\ket{\Psi}$, will have the following form:
	\begin{equation}
		\ket{\Psi} = \sum_{n,n'} A_{n,n'} \ket{n}_{\textsc{c}} \otimes \ket{n'}_{\textsc{r}}. \label{eq:Ann}
	\end{equation}
	Our calculation will proceed in three steps based on this general form.
	\begin{enumerate}
		\item The simplest possibility is a diagonal $A_{n,n'}$. We will demonstrate that this leads, at least according to the PW formalism, to the simplest time evolution possible. We will demonstrate how this time evolution can be interpreted as a particularly boring case of the Novikov self-consistency conjecture.
		\item We derive a quite general statement about $A_{n,n'}$ when it is non-diagonal. Concretely, introducing two sparse matrices $\Delta_1, \Delta_2$ and a diagonal matrix $D$, it will take the form
			\begin{equation}
				A = \Delta_1^T D \Delta_2.
			\end{equation}
		\item A general identity for the conditional probabilities~\eqref{eq:CondProb} in our toy model is derived.
		\item Using the previous two steps, we will prove that the results of the first step carry over to non-diagonal $A_{n,n'}$, too.
	\end{enumerate}
	
	In the \textbf{first step,} let us calculate the components of equation~\eqref{eq:CondProb}. This will tell us straightforwardly what we want to know.
	\begin{align}
		\tr{\HR \hat{P}_{\theta_{\ast}} \hat{\rho}} &= \tr{\sbr{\idC \otimes \HR}\; 
			\sbr{\rbr{\ketasttheta \braasttheta} \otimes \idR}\; \sbr{\ket{\Psi}\negthickspace\bra{\Psi}} },\\
			&= \Bra{{\vphantom{{\sbr{\idC \otimes \HR }}_{\textsc{c}}}}\Psi} \sbr{\rbr{{\ketasttheta}\ {\braasttheta}} \otimes \idR} \; \sbr{\idC \otimes \HR } \Ket{{\vphantom{{\sbr{\idC \otimes \HR }}_{\textsc{c}}}}\Psi}.\label{eq:step1a}
	\end{align}
	With the currently assumed diagonal form for $A_{n,n'}$, we can evaluate
	\begin{equation}
		(\idC \otimes \HR) \ket{\Psi} = \sum_n A_n \rbr{n+\oo{2}} \ket{n}_{\textsc{c}}\otimes \ket{n}_{\textsc{r}},
	\end{equation}
	which in turn—using ${\vphantom{\braket{m|n}}}_{\textsc{r}}\!\braket{m|n}_{\textsc{r}} = \delta_{mn}$—turns equation~\eqref{eq:step1a} into
	\begin{align}
		\tr{\HR \hat{P}_{\theta_{\ast}} \hat{\rho}} &= \sum_n |A_n|^2 \rbr{n+\oo{2}} \abs{{\vphantom{\braket{\psi({\theta}_{\ast})|n}}}_{\textsc{c}}\! \braket{n|\psi(\theta_{\ast})}_{\textsc{c}}}^2.
	\end{align}
	Going back to the definition of $\ketasttheta$, equation~\eqref{eq:ketasttheta}, we note that
	\begin{equation}
		{\vphantom{\braket{\psi({\theta}_{\ast})|n}}}_{\textsc{c}}\! \braket{n|\psi(\theta_{\ast})}_{\textsc{c}} = e^{-i(n+\oo{2}) \theta_{\ast}}\ {\vphantom{\braket{\psi({0}_{\ast})|n}}}_{\textsc{c}}\! \braket{n|\psi(0_{\ast})}_{\textsc{c}}.
	\end{equation}
	This finally leads to
	\begin{equation}
		\abs{{\vphantom{\braket{\psi({\theta}_{\ast})|n}}}_{\textsc{c}}\! \braket{n|\psi(\theta_{\ast})}_{\textsc{c}}}^2 = \abs{{\vphantom{\braket{\psi({0}_{\ast})|n}}}_{\textsc{c}}\! \braket{n|\psi(0_{\ast})}_{\textsc{c}}}^2,
	\end{equation}
	from which we can deduce that
	\begin{equation}
		\tr{\HR \hat{P}_{\theta_{\ast}} \hat{\rho}} = \tr{\HR \hat{P}_{0_{\ast}} \hat{\rho}}, 
	\end{equation}
	and, after a similar lead-up, likewise for the denominator of equation~\eqref{eq:CondProb} that
	\begin{equation}
		\tr{\hat{P}_{\theta_{\ast}} \hat{\rho}} = \tr{\hat{P}_{0_{\ast}} \hat{\rho}}. 
	\end{equation}
	It follows immediately that
	\begin{equation}
		E(\HR|\theta_{\ast}) = E(\HR|0_{\ast}).\label{eq:step1final}
	\end{equation}
	
	Put differently, the result~\eqref{eq:step1final} means that with respect to such diagonal states of the \enquote{universe}, no time will pass, irrespective of the chosen clock states. As was frequently pointed out in the literature, much of the evolution is fixed by the choice of the full state $\ket{\Psi}\in\mathcal{H}$, so at this point this might just be an artefact of an overly specific type of states. Still, even for these states it is interesting that changing clock states cannot change the result. It is important to note that we make no claims concerning more general models; this analysis heavily relies on the structure of our two-harmonic-oscillator model. We will postpone the discussion of the implications to time travel for until after the more general results of \enquote{step three} for non-diagonal $A_{n,n'}$. 
	
	Two examples of states $\ket{\Psi}$ with diagonal $A_{n,n'}$ and isotropic oscillators (\emph{i.e.}, $\omegaC=\omegaR\ifed\omega$) are visualized in figure~\ref{fig:iso}, with different choices for a dispersion parameter $\beta$. The coordinates $a,\chi$ can in this case be assigned to either system (clock or residual), and were chosen to allow easy comparison with the closely related examples of quantum cosmology found in~\cite{Kiefer:1989va}. More details about the construction of the wave packets is relegated to appendix~\ref{app:fig}.
	
	\begin{figure}[t!]
		\centering
		\begin{subfigure}[t]{.49\textwidth}
			\includegraphics[width=\textwidth]{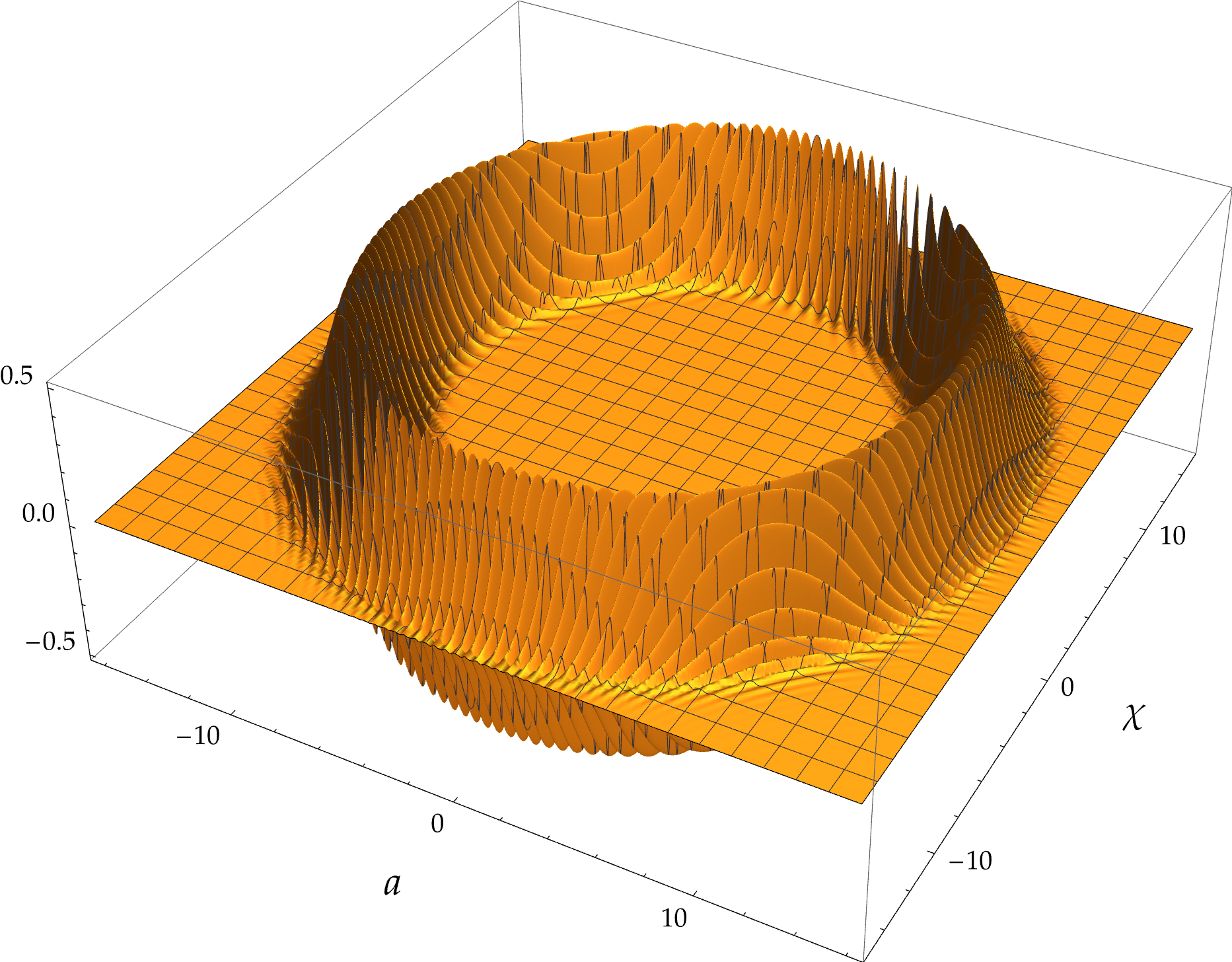}
			\caption{$\beta=\num{1}$}
			\label{fig:iso-no-disp}    
		\end{subfigure}
		\begin{subfigure}[t]{.49\textwidth}
			\includegraphics[width=\textwidth]{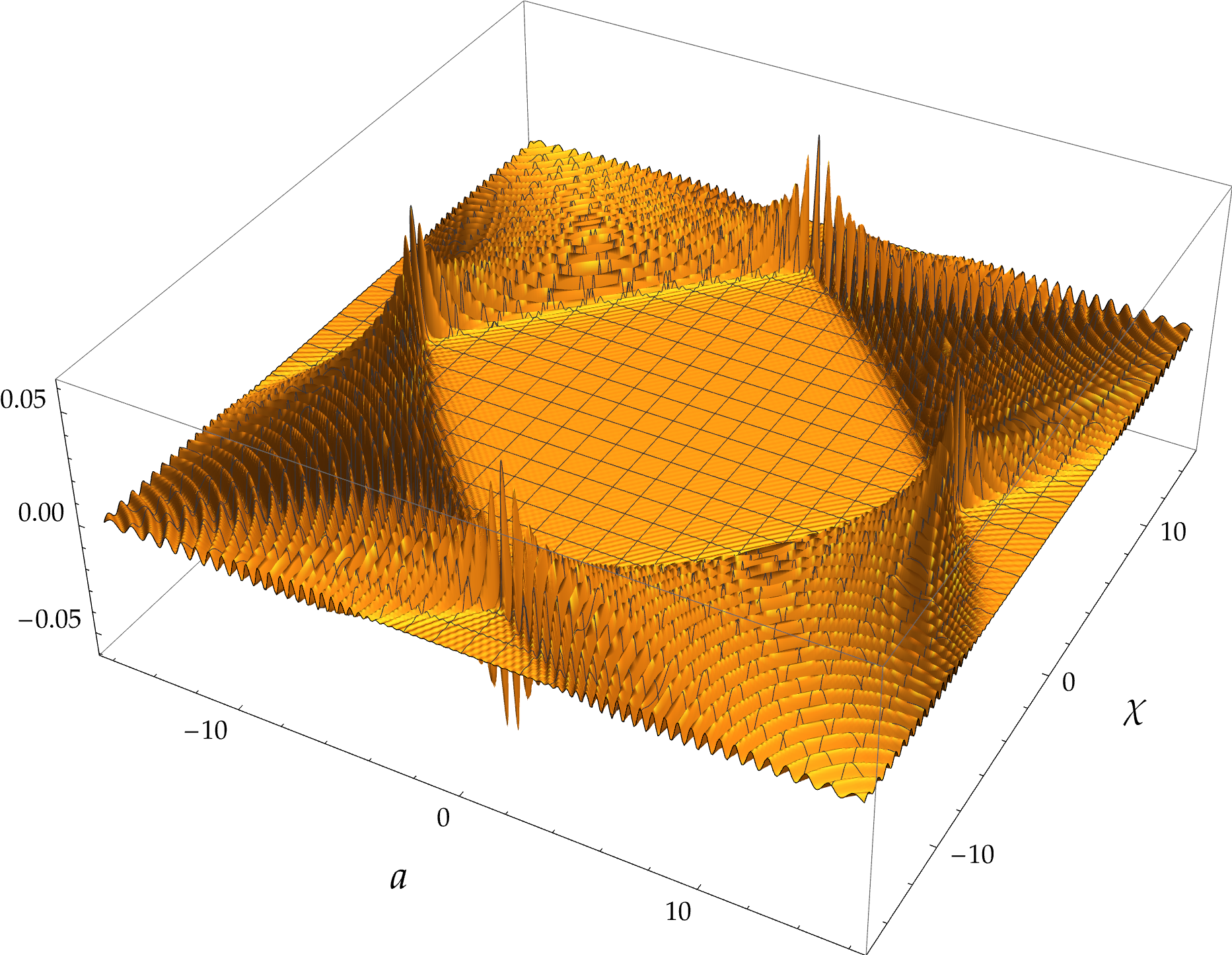}
			\caption{$\beta=\num{0.03}$}
			\label{fig:iso-disp}    
		\end{subfigure}
		\caption{Illustration of wave packets solving our toy model for time travel with $\omegaC=\omegaR$. Details regarding the construction of these solutions to the Hamiltonian constraint equation~\eqref{eq:WdW} can be found in appendix~\ref{app:fig}. In comparison, quantum cosmological applications of this equation constrain themselves to only a half-plane $a\geq0$. More general wave functions with $\omegaC\neq\omegaR$ will have less symmetry than the pictured examples. \textit{Left:} Dispersion-less packets, corresponding to a parameter choice $\beta=1$. \textit{Right:} Wave packets with strong dispersion, $\beta=0.03$.}
		\label{fig:iso}
	\end{figure}
	
	As a preliminary, \textbf{second step} for the discussion of non-diagonal $A_{n,n'}$, it is instructive to reexamine the commensurability condition. Plugging the ansatz~\eqref{eq:Ann} into the Hamiltonian constraint~\eqref{eq:WdW} we get for all values of $n$ and $n'$ that
	\begin{equation}
		\rbr{\omegaC n + \frac{\omegaC}{2} - \omegaR n' - \frac{\omegaR}{2}} A_{n,n'} = 0. 
	\end{equation}
	This implies that either $A_{n,n'}  =0 $ or if $A_{n,n'} \neq 0$  that
	\begin{equation} 
		\rbr{\omegaC n + \frac{\omegaC}{2} - \omegaR n' - \frac{\omegaR}{2}} = 0, 
	\end{equation}
	which is equivalent to the original commensurability condition~\eqref{eq:commensurability}. Given that this condition yields a ratio of two odd numbers, we can rephrase $n$ and $n'$ as
	\begin{equation}
		n = s(2p+1)  +p; \qquad n'=  s(2q+1) +q; \qquad s,p,q\in\mathds{N}_0.
	\end{equation}
	This equation gives additional insight about the coefficients $A_{n,n'}$. In components, we can write
	\begin{equation}
		A_{n,n'} = \sum_{s=0}^\infty \delta_{n, s (2p+1) + p} \; d_s\; \delta_{n', s(2q+1) + q}
	\end{equation}
	which in \enquote{matrix form} appears as follows:
	\begin{equation}
		A =  \Delta_1^T D \Delta_2,\label{eq:step2final}
	\end{equation}
	where $D$ really is diagonal and $\Delta_1, \Delta_2$ are sparse matrices connecting $s$ with $n$ and $n'$, respectively.
	
	The \textbf{third step} takes a look at the numerator of the conditional probability of the PW formalism, equation~\eqref{eq:CondProb}. Note that equation~\eqref{eq:step1a} is fully general, and needed no diagonality of $A_{n,n'}$ in its derivation. Its last two factors can now be evaluated with our general ansatz~\eqref{eq:Ann}:
	\begin{equation}
		(\idC \otimes\HR)\ket{\Psi} = \sum_{n,n'} A_{n,n'} \rbr{n'+\oo{2}} \ket{n}_{\textsc{c}} \otimes \ket{n'}_{\textsc{r}}.
	\end{equation}
	As in the first step, we can then use orthonormality of the Fock states $\ket{n'}_{\textsc{c,r}}$ and cyclicity of the trace to show that
	\begin{equation}
		\tr{\HR \hat{P}_{\theta_{\ast}} \hat{\rho}}=  \sum_{m,n,n'} \rbr{{\vphantom{\braket{m|\psi({\theta}_{\ast})}}}_{\textsc{c}}\! \braket{m|\psi(\theta_{\ast})}_{\textsc{c}} A_{m,n'}^*} \rbr{A_{n,n'} \sbr{n'+\oo{2}}\ {\vphantom{\braket{\psi({\theta}_{\ast})|n}}}_{\textsc{c}}\! \braket{\psi(\theta_{\ast})|n}_{\textsc{c}}}.\label{eq:step3a}
	\end{equation}
	Next, let us employ the definition of $\ketasttheta$, equation~\eqref{eq:ketasttheta}: 
	\begin{align}
		{\vphantom{\braket{m|\psi({\theta}_{\ast})}}}_{\textsc{c}}\! \braket{m|\psi(\theta_{\ast})}_{\textsc{c}}
			&= {\vphantom{\ket{\psi({\theta}_{\ast})}}}_{\textsc{c}}\! \bra{m} e^{-i\HC \theta_{\ast}}\ketastnull,\\
			&= e^{-i(m+\oo{2}) \theta_{\ast}}\ {\vphantom{\braket{m|\psi({0}_{\ast})}}}_{\textsc{c}}\! \braket{m|\psi(0_{\ast})}_{\textsc{c}}.
	\end{align}
	This allows us to rearrange equation~\eqref{eq:step3a} to
	\begin{equation}
		\tr{\HR \hat{P}_{\theta_{\ast}} \hat{\rho}} = \sum_{m,n} Q_{m,n} e^{+i(n-m) \theta_{\ast}}\label{eq:step3b}
	\end{equation}
	where
	\begin{equation}
		Q_{m,n} \defi \rbr{ \sbr{\sum_{n'} A_{m,n'}^* A_{n,n'} (n'+1/2)}\ {\vphantom{\braket{m|\psi({0}_{\ast})}}}_{\textsc{c}}\! \braket{m|\psi(0_{\ast})}_{\textsc{c}}\ {\vphantom{\braket{\psi({0}_{\ast})|n}}}_{\textsc{c}}\! \braket{\psi(0_{\ast})|n}_{\textsc{c}}}\label{eq:step3c}
	\end{equation}
	This effectively demonstrates how the choice of the universe's state $\ket{\Psi}$ and the chosen initial clock state $\ketastnull$ together fully determine the time evolution of our model. Equation~\eqref{eq:step3c} can equivalenty be expressed in \enquote{matrix form} as
	\begin{equation}
		Q = D_1^\dagger A^\dagger D_0 A D_1,\label{eq:step3final}
	\end{equation}
	where $D_1$ is a complex diagonal \enquote{matrix}, and $D_0$ a real diagonal \enquote{matrix}. 
	
	A closely related identity can also be found—using similar arguments—for the denominator of equation~\eqref{eq:CondProb}, reading: 
	\begin{equation}
		\tr{ \hat{P}_{\theta_{\ast}} \hat{\rho}}=  \sum_{m,n} \rbr{ \sbr{ \sum_{n'} A_{m,n'}^* A_{n,n'}}\;	{\vphantom{\braket{m|\psi({{0}_{\ast}})}}}_{\textsc{c}}\!\braket{m|\psi(0_\ast)}_{\textsc{c}}\ {\vphantom{\braket{\psi({0}_{\ast})|n}}}_{\textsc{c}}\! \braket{\psi(0_{\ast})|n}_{\textsc{c}} e^{+i(n-m) \theta_{\ast}}}.\label{eq:step3final2}
	\end{equation}
	Note that this is again of the form~\eqref{eq:step3b}, though with a simpler \enquote{$Q$} for which $D_0=\mathds{1}$. This means any general statement that can be shown for $Q$ will hold for the corresponding part appearing in equation~\eqref{eq:step3final2}.
	
	In our \textbf{fourth and final step,} we shall bring together our result~\eqref{eq:step2final} for $A_{n,n'}$ and the result~\eqref{eq:step3final} for $Q_{m,n}$ to greatly constrain the conditional probabilities of equation~\eqref{eq:CondProb}. Concretely, inserting the former into the latter gives:
	\begin{align}
		 Q 	&= D_1^\dagger (\Delta_1^T D \Delta_2)^\dagger D_0 (\Delta_1^T D \Delta_2) D_1,\\
		 	&= D_1^\dagger \Delta_2^\dagger D^* (\Delta_1^* D_0 \Delta_1^T) D \Delta_2 D_1.\label{eq:step4a}
	\end{align}
	In this expression, let us investigate the central, bracketed term, $\Delta_1^* D_0 \Delta_1^T$: 
	\begin{align}
		\rbr{\Delta_1^* D_0 \Delta_1^T}_{ss'} &= \sum_{n'} \delta_{n', s(2q+1) +q} [D_0]_{n'} \delta_{n', s'(2q+1) +q},\label{eq:step4b}\\
			&\ifed (D_{3/2})_{ss'},
	\end{align}
	another diagonal matrix. This in turn means that even more terms in the middle of equation~\eqref{eq:step4a} reduce to yet another diagonal matrix $D_2$, defined as
	\begin{equation}
		D_2 \defi D^* (\Delta_1^* D_0 \Delta_1^T) D = D^* D_{3/2} D.
	\end{equation}
	This, then, allows us to greatly simplify the structure of $Q$ itself:
	\begin{equation}
		Q= D_1^\dagger (\Delta_2^\dagger D_{2} \Delta_2) D_1,
	\end{equation}
	whose middle part in brackets can, using the same approach as in equation~\eqref{eq:step4b}, be shown to be diagonal. Finally, it follows that $Q$ itself is diagonal.
	
	Essential for this result was the specific form of $A$, equation~\eqref{eq:step2final}. Modifying the underlying model will therefore likely change this result. For the PW formalism in our context, though, the result is far-reaching: If we take equation~\eqref{eq:step3b}, we see that for diagonal $Q$, the term $e^{i(n-m)\theta_{\ast}}=1$, as $n=m$. The same observation holds for the denominator of the PW conditional probabilities~\eqref{eq:CondProb}, equation~\eqref{eq:step3final2}. So, as in our first step when considering diagonal $A_{n,n'}$, even the general case produces trivial time evolution.
	
	At first glance, this seems a very unremarkable result, and therefore of little use for any interesting physics question. Ironically, this is \emph{not} the case in our present situation, in which we are discussing time travel: An evolution that is simply constant is certainly still trivially a case of Novikov's self-consistency conjecture. Given that nothing is happening, one might even go so far as to say that it gives an odd entry to the intersection of the Novikov self-consistency conjecture and the \enquote{boring physics conjecture}~\cite{Visser:1995cc}. It is simultaneously both time travel and the absence of all evolution.
	
	This result is also a marked departure from results in quantum cosmology~\cite{Conradi:1990rw,Brunetti:2009eq}. The reason for this disparity is the limited domain of the scale parameter in the models of the form~\eqref{eq:WdW}. Most importantly, limiting a harmonic oscillator to half of its configuration space will prevent the application of ladder operators as we used them in the above analysis. This again demonstrates how \enquote{simple} changes to the model can quite radically change results.

	\section{Interpretation, Discussion and Outlook}\label{sec:final}
	Despite the disappointingly boring outcome for our particular model, our toy model serves its purpose as a proof-of-concept study. It demonstrates the feasibility for future model building of quantum systems with periodic, emergent notions of time as a tool to study time travel. The obvious model-dependence of the PW formalism through its dependence on $\ket{\Psi}$, the \enquote{state of the universe}, means that different models might have very different conclusions as to what sort of time evolution is possible. As even collections of harmonic oscillators become increasingly complex once interactions start to enter, necessitating many different heuristics or approaches across fields~\cite{AulettaFortunatoParisi2009QM,Zee:2010qce}, it is unlikely that the PW formalism can be fully general even when discussing whether time travel is possible based on first principles of quantum theory.
	
	In the absence of a fully general result, we still believe that some general conclusions can nevertheless be achieved. Analytically tractable models beyond the one presented in our study should be possible, especially given that POVMs and the PW formalism feature prominently in the modern, gauge theoretic consolidation of emergent time in quantum theory. The use of POVMs in future extensions was part of the motivation to highlight them as much in our discussion as we did, despite the final result not relying on them. 

	The goals of such a future model building task would contain the following:
	\begin{itemize}
		\item More and different subsystems; more in the sense of more harmonic oscillators, different in adding various combinations of other systems. Also ancillary systems could be used. For example, in the context of time travel, one can ask the following question: Can ancillary or memory systems allow for (more or less self-consistent) time travel being limited to just a fixed, finite number of round trips as opposed to the infinite traversability of GR's CTCs?
		\item In a closely related step, one would introduce various interactions among the subsystems. These will greatly change the outcome of our above analysis.
		\item Ultimately, the combination of these two model building drives can try to investigate entropic arguments in the context of time travel. If self-consistent time travel in quantum systems with an emergent notion of time is possible, are such configurations potentially favoured or disfavoured on thermodynamic grounds? In particular, entropy concepts geared towards constrained or fixed energy systems, like \enquote{observational entropy}~\cite{SafranekDeutschAguirre2019QuCoarseGrEntr,SafranekDeutschAguirre2019QuCoarseGrEntrClosed,Safranek:2020tgg}, promise access to new arguments for or against time travel without having to rely on either mixing different concepts of time (GR's dynamic time \emph{vs.} thermodynamic time \emph{vs.} parameter time \emph{vs.} \dots) or, in fact, without relying on any kind of explicit notion of time; the entropic arguments in such systems would similarly be based on emergent notions, just as the notion of time itself is. 
		\item Similarly, this exercise should be able to further clarify the difference between a periodic clock time and time travel as such. Obviously, some kind of synchronicity between residual system and clock system has to hold for the latter. Yet the above mentioned research directions would allow one to turn concepts like self-consistency into less of a binary. After all, if time is emergent, its absence might be ill-suited for human survival but still allow a valid physical system that describes something \enquote{close} to a quantum system with self-consistent, observable time travel. In a sense, this would be a first step towards rigorously testing Hawking's chronology protection. 
		\item Much of the current research which led to a reimagining of the PW formalism as a gauge fixed picture aims to ask what happens when one changes frames, \emph{i.e.}, clocks. Once the model systems for time travel are complex enough (and, thus, more complex than our equation~\eqref{eq:WdW}), a natural question is: What happens to time travel in such quantum reference frame transformations? And, linking to the previous point, does this differ from the situation for merely periodic clocks?
		\item Lastly, the previous two points hint at the possibility of making time travel more a local, emergent notion in a large system. This is, for one, the explicit way to study time travel classically as CTCs in a space-time manifold, see, for example,~\cite{Earman:2009mda,Dolansky:2010nr}. For another, this is similarly done in quantum physics, ranging from the rigorously and explicitly local~\cite{Bishop:2020qtt,Alonso-Serrano:2021ydi} to the more imprecise and only implicitly local~\cite{Deutsch:1991nm,Lloyd:2011zz}. Yet, so far, these approaches all had to rely on an \emph{ad-hoc} introduction of \enquote{background} time.
	\end{itemize} 
	
	There are other avenues along similar lines that warrant a closer look in the framework of quantum gravity. In the canonical approach to it, for instance, different inner products can be considered, so at least changing the mathematical framework drastically—begging the question how much of the physical interpretation changes or has to change in tandem. Other approaches to quantum gravity besides the canonical ones can also test emergent notions of time in their respective frameworks. Besides a question of tractability, this was part of the motivation to keep our toy model somewhat agnostic and conservative with respect to a theory of quantum gravity.
	
	Independent of which direction one would want to take, we believe that our first result shows that emergent notions of time in quantum physics can lead to a multitude of questions and interpretational conundra pertaining to time travel. The very introduction of such an emergent time concepts undermines most traditional arguments against time travel. The concepts evoked in these arguments simply have to rely on notions that are not valid in more \emph{ab initio} approaches.
	
	\funding{A.A.-S. is funded by the Deutsche Forschungsgemeinschaft (DFG, German Research Foundation) — Project ID 516730869. This work was also partially supported by Spanish Project No. MICINN PID2020-118159GB-C44. S.S. was financially supported by Czech Science Foundation grant GACR~23-07457S. M.V. was directly supported by the Marsden Fund, via a grant administered by the Royal Society of New Zealand.}
	
	\dataavailability{No new data were created or analyzed in this study. Data sharing is not applicable to this article.} 
	
	\acknowledgments{S.S. acknowledges support from the technical and administrative staff at the Charles University. He thanks Emily Adlam, Antonis Antoniou, Julian Barbour, Martin Bojowald, Leonardo Chataignier, Fabio Costa, Ricardo Faleiro, Klaus Fredenhagen, Philipp A. Höhn, Claus Kiefer, Jorma Louko, Jessica Santiago, Alexander R. H. Smith, Reinhard Werner, Magdalena Zych, and the participants of the 781\textsuperscript{st} WE-Heraeus-Seminar \enquote{Time and Clocks} in Bad Honnef, the Geometric Foundations of Gravity 2023 conference in Tartu, the 13\textsuperscript{th} RQI-North meeting in Chania, and the Odborné soustředění ÚTF in Světlá pod Blaníkem for many valuable discussions and their interest.}
	
	\conflictsofinterest{The authors declare no conflict of interest. The funders had no rôle in the design of the study; in the collection, analyses, or interpretation of data; in the writing of the manuscript; or in the decision to publish the results.} 
	
	
	\abbreviations{Abbreviations}{
		The following abbreviations are used in this paper:\\
		
		\noindent 
		\begin{tabular}{@{}ll}
			CTC		& Closed, time-like curves\\ 
			GR		& General relativity\\
			POVM	& Positive, operator-valued measure\\
			PVM		& Projection-valued measure\\
			PW 		& Page--Wootters\\
			WDW		& Wheeler--DeWitt
		\end{tabular}
	}
	
	\appendixtitles{yes} 
	\appendixstart
	\appendix
	\section[\appendixname~\thesection]{Formal Definition of POVMs}\label{app:defPOVM}
	Let us present in this appendix the formal definition of a POVM. An operator $\hat{A}$ on a Hilbert space $\mathcal{H}$ is called \emph{positive}, if 
	\begin{equation}
		\forall\ket{\psi}\in \mathcal{H}:\qquad \braket{\psi|\hat{A}|\psi} \geq 0. 
	\end{equation} 
	Positivity of an operator is often denoted as \enquote{$\hat{A}\geq0$.}	An operator $B$ is called a \emph{POVM} if and only if it is an \emph{additive} map from a Borel $\sigma$-algebra $\mathcal{A}$ of subsets of a set $\Omega$ to operators on $\mathcal{H}$ such that
	\begin{itemize}
		\item[(i)] for all sets $X\in \mathcal{A}$, $B(X)\geq 0$ (\enquote{positivity}),
		\item[(ii)] for disjoint $X,Y\in \mathcal{A}$, $B(X\cup Y) = B(X) + B(Y)$ (\enquote{additivity}),
		\item[(iii)] for all sequences of sets $X_n \in \mathcal{A}$ such that $\cap_{n=1}^\infty X_n = \emptyset$, all operators $B(X_n)$ have $\hat{0}$ (the zero operator) as greatest lower bound.
	\end{itemize}
	For more details, see \cite[p.71, Def.4.5]{BLPY2016QuantumMeasurement}. A POVM that fulfils $B(\Omega)=\mathds{1}$ is called \emph{normalized}.
	
	Normalized POVMs are a more general notion of observable compared to that of observables being unbounded, self-adjoint operators on $\mathcal{H}$. The (set-theoretic) \enquote{universe} $\Omega$ corresponds to possible measurement outcomes. In this sense, additivity and being normalized links measurements as described by POVMs to the theory of probablity as set down in the Kolmogorov axioms. They allow for more and wider notions of imprecision than those encoded in density matrices~\cite{AulettaFortunatoParisi2009QM,BLPY2016QuantumMeasurement}. In \enquote{standard quantum mechanics,} The measurement process of an observable $\hat{A}$ according to the Born rule projects onto a subspace formed by eigenvectors of $\hat{A}$ with set eigenvalue $A_i$. In the terminology of POVMs this corresponds to \enquote{projection-valued measures} (PVMs), and PVMs $\subsetneq$ POVMs. 
 
	\section[\appendixname~\thesection]{Time Operators for a Harmonic Oscillator}\label{app:HO}
	In this appendix, we want to draw attention to some additional background regarding POVMs and phase or time operators in the context of harmonic oscillators.
	
	With the POVM $B_0$ of equation~\eqref{eq:B0} in place, we can now define a large number of new, self-adjoint operators~\cite{BGL1994TimeObservables}. To do this, we associate to each real-valued, bounded, and measurable function $f$ on $[0,2\pi)$ the operator $\hat{B}_0[f]$ defined as
	\begin{align}
		\hat{B}_0[f] &\defi \oo{2\pi} \int_0^{2\pi} f(\theta) \ket{\theta}\negthickspace\bra{\theta} \dif \theta,\\
		&= \sum_{n,m\geq 0} \oo{2\pi} \int_0^{2\pi} e^{i(n-m)\theta} f(\theta) \ket{n}\negthickspace\bra{m} \dif \theta.
	\end{align}
	This procedure fixes the set $X$ in the original POVM $B_0$, and then uses it to construct various Toeplitz operators. Different choices for $f$ can be found in the literature (see, for example,~\cite{Garrison:1970}), but we will focus on one of the simplest options:
	\begin{equation}
		f(\theta) = \theta.
	\end{equation}
	This will give us a time operator $\hat{T}_0$, which we calculate with the above expressions to be
	\begin{equation}
		\hat{T}_0 = \hat{B}_0[\theta] = \sum_{n\neq m \geq 0} \oo{i(n-m)} \ket{n}\negthickspace\bra{m} + \pi \mathds{1}.
	\end{equation}
	
	However, this operator is not unique. First of all, we could have chosen a different starting point for our angles, changing the index $0$ at that point. There is also a different, but more physically motivated point of arriving at additional time operators. The time operator $\hat{T}_0$ can be covariantly shifted to a new operator $\hat{T}_{\theta_{\ast}}$:
	\begin{equation}
		\hat{T}_{\theta_{\ast}} \defi e^{i\hat{H}\theta_{\ast}}  \hat{T}_0  e^{-i\hat{H}\theta_{\ast}}.
	\end{equation}
	This \emph{covariant shift property} is a general feature of such time operators from POVMs, see~\cite{BGL1994TimeObservables}, that is increasingly prominent and clear in their modern implementations~\cite{Hoehn:2019fsy,Smith2019GTwirl}. Importantly, while all such $\hat{T}_{\theta_{\ast}}$ fulfil canonical commutation relations with $\hat{H}$ (when the resulting operator is well-defined), different time operators will not commute with each other:
	\begin{equation}
		\theta_{\ast}\neq \theta_{\ast}' \mod 2\pi \qquad \Longrightarrow \qquad [\hat{T}_{\theta_{\ast}} \hat{T}_{\theta_{\ast}'}] \neq 0.
	\end{equation}
	
	These results deserve some commentary. The reason why these time operators are still self-adjoint, despite the illustrious list of no-go theorems given in the introduction, is two-fold~\cite{BGL1994TimeObservables}: One, they are hardly unique. Any change in starting point (phase) will yield a different time operator. A covariant shift similarly yields new time operators. Two, the domain of time is now periodic. The no-go theorems usually aim for $\mathds{R}$ as domain for time $\theta$, or that the time would be unique in some sense. In the present context, it is more a question of when one started the periodic clock, and what period the clock has. It is also worth pointing out that our time operators subtly fail to satisfy the expected time-energy uncertainty: It will not hold for all states in $\mathcal{H}$, but only for a densely defined set~\cite{Garrison:1970,Galindo:1984}. This latter restriction is not an issue for our purposes, at least in the present article, and the other reasons for evading the no-go theorems are what allows our application in the context of time travel.
	
	\section[\appendixname~\thesection]{Detailed Construction of Depicted Wave Packets}\label{app:fig}
	For our figure~\ref{fig:iso}, we adapted the plots of~\cite{Kiefer:1989va} to our present context with a larger configuration space. For ease of comparison, we kept the notation for the wave functions the same. We do, however, emphasize that in our context one should \emph{not} interpret $a$ as a scale factor. Our investigation is not quantum cosmology, just inspired by a particular model of it due to its simplicity.
	
	A general solution of equation~\eqref{eq:WdW} can be expressed in configuration space variables $a,\chi$ instead of the Fock space picture adapted in the remainder of the article. For illustration purposes, this is preferable. We will restrict attention to symmetric wavefunctions, in which $a$ and $\chi$ can be assigned equally well to the clock system $\HC$ or the residual system $\HR$. While our enlarged configuration space (compared to that of quantum cosmology) might change the overall normalization, it will not change the qualitative behaviour of the wavefunctions. We therefore also copy the normalization of reference~\cite{Kiefer:1989va}.
	
	A wavefunction with two Gaussians centered around $\chi=\pm \chi_0$, each of width $\beta$, at time/position $a=0$, is given by
	\begin{equation}
		\Psi(a,\chi) = \sum_{m=0}^\infty c_{2m} \frac{H_{2m}(a) H_{2m}(\chi)}{\sqrt{2^{2m} (2m)!} H_{2m}(0)} \exp\rbr{-\oo{2} a^2 - \oo{2} \chi^2},\label{eq:appPsi}
	\end{equation}
	where $H_n(x)$ is the $n$-th Hermite polynomial and the expansion coefficients are
	\begin{equation}
		c_n = \oo{\sqrt{2^n n!}} \sqrt{\frac{2\beta}{1+\beta^2}} \exp\rbr{- \frac{\chi_0^2}{2(1+\beta^2)}}\rbr{\sqrt{\frac{1-\beta^2}{1+\beta^2}}}^n H_n \rbr{\frac{\chi_0}{\sqrt{1-\beta^4}}}.
	\end{equation} 
	These coefficients $c_n$ simplify further for the \enquote{coherent case}, \emph{i.e.}, when $\beta=1$, to
	\begin{equation}
		c_n = \exp\rbr{-\oo{4}\chi_0^2}\frac{\chi_0^n}{\sqrt{2^n n!}}.
	\end{equation} 
	Note that the index $m$ in equation~\eqref{eq:appPsi} is meant to make more explicit that the sum runs only over even $n$.
	\begin{adjustwidth}{-\extralength}{0cm}
		
		\reftitle{References}
		
		
		
	\end{adjustwidth}
\end{document}